\documentclass[review,3p]{elsarticle}
\usepackage{graphicx}
\usepackage{amsthm}
\usepackage{amssymb,color}
\usepackage{bm,amsmath,longtable,booktabs, array,rotfloat}
\usepackage{subfigure}

\usepackage{nomencl,multicol}

\journal{Elseiver Science}

\begin{document}

\begin{frontmatter}



\title{Lattice Boltzmann simulation of separation phenomenon in a binary gaseous flow through a microchannel}

%
\author[label1,label2]{Liang Wang}
\address[label1]{State Key Laboratory of Coal Combustion, Huazhong University of Science and Technology, Wuhan, 430074, China}
\address[label2]{Beijing Computational Science Research Center, Beijing 100193, China}
\author[label1]{Zimian Xu}
\author[label1,label2]{Zhaoli Guo\corref{cor1}}
\ead{zlguo@hust.edu.cn}
\cortext[cor1]{Corresponding author.}

\begin{abstract}
Gas separation of a binary gaseous mixture is one of characteristic phenomena in the micro-scale flows that differ from the conventional size flows. In this work, the separation in a binary gas mixture flows through a microchannel is investigated by the lattice Boltzmann method with a diffuse-bounce-back (DBB) boundary condition. The separation degree and rate are measured in the He--Ar and Ne--Ar systems for different mole fractions, pressure ratios, and Knudsen numbers. The results show that the separation phenomenon in the He--Ar mixture is more obvious than that in the Ne--Ar mixture at the same mole fraction owing to the larger molecular mass ratio. In addition, the increase in the pressure ratio reduces the difference in the molecular velocities between the two species, and the separation phenomenon becomes weaker. However, the gas separation is enhanced with an increase in the Knudsen number. This is because the resulting rarefaction effect reduces the interactions between the gas molecules of the two species, and thus increases the difference in the molecular velocity.
\end{abstract}


\end{frontmatter}


\makenomenclature
\begin{multicols}{2}
\nomenclature{$C_{l}$}{Mole fraction of the light species}
\nomenclature{$P$}{Pressure}
\nomenclature{$k_B$}{The Boltzmann constant}
\nomenclature{$T$}{Temperature}
\nomenclature{$\theta$}{Ratio of the inlet and outlet pressure}
\nomenclature{$\lambda$}{Mean free path}
\nomenclature{$CL$}{Separation degree}
\nomenclature{$\sigma$}{a or b species in gas mixture}
\nomenclature{$n_{\sigma}$}{Number densities}
\nomenclature{$n$}{$\sum_{\sigma}n_{\sigma}$}
\nomenclature{$x_{\sigma}$}{$n_{\sigma}/n$}
\nomenclature{$\mu_{\sigma}$}{viscosities of the $\sigma$ species}
\nomenclature{$m_{\sigma}$}{Molecular weight}
\nomenclature{$m_{min}$}{$min(m_{a}, m_{b})$}
\nomenclature{$\mathrm{Kn_{out}}$}{Knudsen number at the outlet}
\nomenclature{$N$}{Number of lattice grid in the characterize length}
\nomenclature{$H$}{Channel height}
\nomenclature{$L$}{Channel length}
\printnomenclature

\end{multicols}

\section{\label{intro}Introduction}

With the rapid development of vacuum technology and micro electro mechanical systems (MEMS) as well as energy converters such as solid oxide fuel cells (SOFCs), considerable interests have been attached to the rarefied gas mixture flows \cite{Karniadakis,Squires,Nie,Chapman1970,Zhang2011,Wu}. Due to the microscopic interactions between the gas molecules from different species besides those from the same species, the gas mixture flows are more complex than the single-species gas flows. It has been shown that some special phenomena in the gas mixture flows can appear compared with the single gas flows \cite{Present,Szalmas,Szalmas2,Myong,Dodulad}. As one of these phenomena, the gas separation has been received much attention, and is attributed to the difference in the mean velocities of each species \cite{Sharipov05}.

The separation phenomenon in gas mixtures is very important in many applications including pumping, sampling, filtering, etc. \cite{Sharipov05,Takata07,Varoutis09}. For example, this kind of separation effect should be considered in predictions of the mass flow rate of mixture as well as the flow rate of gas components. Early in 1949, the separation phenomenon was reported first by Present {\sl et al.} ~\cite{Present} who studied the binary gaseous flows through a long circular capillary. They pointed out that the pressure gradient may lead to a diffusion in a mixture flows. With a finite rarefaction effect, different species of the mixtures travel with different speeds in the channel owing to their different molecular velocities, and hence the species tend to separate. Moreover, it is found that the maximum separation degree depends on the molecular mass ratio. On the basis of this theory, some researches have been further performed on the gas separation effect \cite{Dodulad,Vargo,McNamara,Aoki}. Higashi et al. \cite{Higashi64} applied a derived expression of surface diffusion coefficient to predict the degree of separation in binary gas-mixtures, and provided theoretical and experimental verifications through the separation results of the n-butane and propane mixture flows. Via a detailed investigation on the separation phenomenon for binary mixture flow through a long tube into a vacuum, Sharipov et al. \cite{Sharipov05} found that the concentration and rarefaction at the entrance of the tube have a significant effect on the concentration varying along the tube. Dodulad {\sl et al.} \cite{Dodulad} studied the gas separation in a Knudsen pump by solving the Boltzmann equation and demonstrated the influence of molecular potential on the gas separation rate. By solving the linearized Boltzmann equation, Kalempa {\sl et al.} \cite{Kalempaa} also investigated the separation phenomenon and found that the flow rate of each species is influenced by both the mole fraction and rarefaction effect. Szalmas et al. \cite{Szalmas10} employed the McCormack kinetic model to simulate the flows of binary gaseous mixtures through micro-channels with triangular and trapezoidal configurations, and reported that the mass flow rate will be predicted with a 10\% discrepancy when the separation effect is not considered.

From the available studies in the literature, it is indicated to us that the molecular mass ratio and mole fraction have a significant effect on the separation in the binary gaseous flows. However, it is not clear how these internal parameters influence the separation phenomenon. In general, the separation phenomenon in gaseous mixture flows is also determined by some important external factors, such as the compression effect and rarefaction effect. Therefore, a detailed investigation is highly desired for us to reveal the influence of such parameters on the gas-gas separation. As far as we know, no previous works have been reported on this subject. In this work, we will conduct a parametric study to understand how these mentioned parameters affect the separation phenomenon in binary gaseous mixtures, which can be served as a typical case for the separation effect of multicomponent gas mixture systems.

In most works on the separation phenomenon in gaseous mixtures flows, the Boltzmann equation is needed to be trivially solved. As a mesoscopic method which stems from a discrete approximation to the Boltzmann equation, the lattice Boltzmann equation (LBE) method has been recently applied to microflows of binary mixtures owing to its simplicity and efficiency ~\cite{Arcidiac07,Joshi07,Szal08,Guo2009}. However, the gas separation of binary mixtures is not investigated in these works. On the other hand, as revealed in the LBE method literature \cite{Guo2009,Lim2002}, the multiple-relaxation-time (MRT) collision model has a better potential for simulating micro-gaseous binary mixtures. Therefore, in this work the binary gaseous flows are simulated by the MRT-LBE model combined with the effective relaxation times and a proposed diffusive-bounce-back boundary (DBB) scheme. The effects of four parameters, including the pressure ratio, the Knudsen (Kn) number, the molecular mass and mole fraction, are investigated on the separation of binary gaseous flow in a microchannel.

The remainder of this paper is organized as follows. In Sec. \ref{problem}, the problem investigated in this work is briefly described, and the MRT-LBE model and the boundary condition for studying the binary micro-gas flows are introduced in Sec. \ref{Model}. In Sec. \ref{annlyze}, numerical results and some discussions are provided, and the conclusions are given in Sec. \ref{results}.

\section{Problem description}\label{problem}
\begin{figure}
\centering
\includegraphics[scale=0.8]{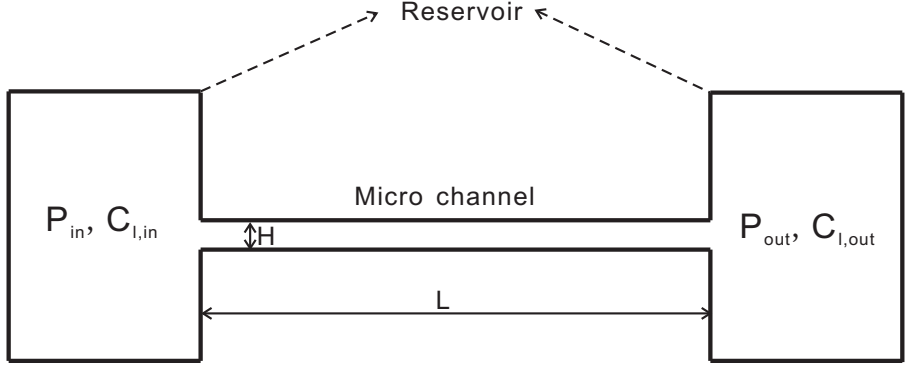}
\caption{Pressure-driven gaseous flows of binary mixtures through a microchannel.}
\label{channel}
\end{figure}

In this work, we consider the two-dimensional flow of gas mixtures in a microchannel of length $L$ and height $H$ as shown in Fig.~\ref{channel}.
The channel connects two reservoirs which are filled with the binary mixture with different pressures and concentrations. The concentration and pressure of the mixture in the reservoirs are assumed to be constant during the flow process since the number of gas molecules flowing through the channel is negligible compared to the number of gas molecules in the reservoirs. In this work we simulate the flow in the microchannel only, and replace the reservoirs with boundary conditions at the inlet and outlet of the channel. Moreover, we set $L/H=35$ in our study, meaning that the end effect at the inlet and outlet can be ignored at this ratio~ \cite{Xu2013}.

In general, the flow described above is influenced by a number of parameters. In this study, we mainly consider the pressure ratio, Knudsen number, molecular mass ratio, and mole fraction of each species. Two binary mixtures are considered here. One is the He--Ar mixture with a molecular mass ratio of $1:10$, and the other is the Ne--Ar mixture with a molecular mass ratio of $1:2$. In these mixtures, the mole fraction of the light species is defined by
\begin{equation}\label{C}
C_{l}=\frac{n_{l}}{n_{l}+n_{h}},
\end{equation}
where $n_{l}$ and $n_{h}$ are the molar densities of the light and heavy species, respectively. The species pressure of the mixture is given by the equation of state:
\begin{equation}\label{P}
   P=n k_B T,
\end{equation}
where $n$ is the molar densities of the gas mixture, and $k_B$ is the Boltzmann constant. The pressure of gas mixture and concentration of the lighter species in the gas mixtures at the inlet and outlet are defined as ($P_{in}$, $C_{l, in}$) and ($P_{out}$, $C_{l, out}$), respectively. The flow is driven by an imposed pressure gradient ($P_{in}>P_{out}$), and the pressure ratio is defined as $\theta=P_{in}/P_{out}$.

Table \ref{C3table1} lists the considered cases, and a $20 \times 700$ lattice is employed in the simulations. $\textrm{Kn}_{out}$ is the given Knudsen number at the outlet and defined by the mean-free path $\lambda_{out}$ of the gas mixture at the outlet, i.e., $\textrm{Kn}_{out}=\lambda_{out}/H$.
\begin{table}[htbf]
\centering
\caption{The simulation cases for the He/Ar and Ne/Ar gas mixtures.}\label{C3table1}
\vspace{6pt}
\begin{tabular}{|c|c|c|c|c|}
\multicolumn{5}{c}
  {\bfseries He/Ar gas mixture}\\ \hline
  Case & $C_{l,in}$ & $C_{l,out}$ & $\textrm{Kn}_{out}$ & $\theta$ \\ \hline
 1 & 30\% & 30\% &  0.3 & 4.0 \\
 2 & 50\% & 50\% &  0.3, 0.5, 1.0, 3.0, 5.0 & 2.0, 3.0, 4.0, 5.0 \\
 3 & 70\% & 70\% &  0.3 & 4.0  \\
 4 & 70\% & 30\% &  0.3, 1 & 2.0 \\
 5 & 30\% & 70\% &  0.3, 1 & 2.0  \\
\hline
 \multicolumn{5}{c}
  {\bfseries Ne/Ar gas mixture}\\ \hline
  Case & $C_{l,in}$ & $C_{l,out}$ & $\textrm{Kn}_{out}$ & $\theta$ \\ \hline
 6 & 30\% & 30\% &  0.3 & 4.0 \\
 7 & 50\% & 50\% &  0.3, 0.5, 1.0, 3.0, 5.0 & 2.0, 3.0, 4.0, 5.0 \\
 8 & 70\% & 70\% &  0.3 & 4.0 \\
 9 & 70\% & 30\% &  0.3, 1 & 2.0\\
 10 & 30\% & 70\% &  0.3, 1 & 2.0  \\
\hline
\end{tabular}
\end{table}
In order to characterize the degree of gas separation, we define the parameter $CL$ as follows:
\begin{equation}\label{CL}
CL=\frac{C_{l, in}-C_{l, min}}{C_{l, in}},
\end{equation}
where $C_{l, min}$ is the smallest mole fraction of the light species in the channel. And another parameter $\widetilde{C_{l}}$ is also needed to represent the variation tendency of total concentration, which is defined as the average quantity of the separation degree:
\begin{equation}\label{Clline}
\widetilde{C_{l}}=\frac{\sum\limits_{x}(C_{l, x}-C_{l, in})}{N_{x} C_{l, in}},
\end{equation}
where $N_{x}$ is the number of lattice grid in the flow direction.

\section{Numerical method and its validation} \label{Model}
It has been shown in the kinetic theory that the LBE method can be directly derived from the continuous Boltzmann equation \cite{He97,Abe97}. In this section, the MRT-LBE model with effective relaxation times and the kinetic boundary condition for of a microscale binary mixture are proposed and subsequently validated.

\subsection{LBE Model for binary gas mixtures}
\label{method}
In this work, a two-dimensional nine-velocity (D2Q9) MRT-LBE model \cite{Guo2009} is used. The evolution of the distribution function is expressed as follows
\begin{equation}\label{MRT}
f_{\sigma i}(\bm{x}+\bm{c}_{i}\delta_t, t+\delta_t)=f_{\sigma i}(\bm{x},t)+\Omega_{\sigma i}(f),~~~i=0,1,\cdots,8,
\end{equation}
where $f_{\sigma i}(\bm{x},t)$ is the distribution function for species $\sigma (a~and~b)$ associated with the gas molecules moving with the discrete velocity $\bm{c}_{i}$ at position $\bm{x}$ and time $t$, $\Omega_{\sigma i}(f)$ is the discrete collision operator defined by
\begin{equation}\label{MR}
\Omega_{\sigma i}(f)=\sum_j(\bm{M}^{-1}\bm{S}\bm{M})_{ij}\left[f_{\sigma j}-f_{\sigma j}^{eq}\right],
\end{equation}
where $\bm{M}$ is a $9\times9$ transform matrix projecting $f_{\sigma i}$ onto the moment space
\begin{equation}\label{m}
\bm{M}=
  \left(
  \begin{array}{rrrrrrrrr}
    1 & 1 & 1 & 1 & 1 & 1 & 1 & 1 & 1\\
    -4 & -1 & -1 & -1 & -1 & 2 & 2 & 2 & 2\\
    4 & -2 & -2 & -2 & -2 & 1 & 1 & 1 & 1\\
    0 & 1 & 0 & -1 & 0 & 1 & -1 & -1 & 1\\
    0 & -2 & 0 & 2 & 0 & 1 & -1 & -1 & 1 \\
    0 & 0 & 1 & 0 & -1 & 1 & 1 & -1 & -1 \\
    0 & 0 & -2 & 0 & 2 & 1 & 1 & -1 & -1\\
    0 & 1 & -1 & 1 & -1 & 0 & 0 & 0 & 0 \\
    0 & 0 & 0 & 0 & 0 & 1 & -1 & 1 & -1\\
  \end{array}
  \right),
\end{equation}
such that $\bm{m}_{\sigma}=\bm{M}\bm{f}_{\sigma}$, where $\bm{f}_{\sigma}=(f_{\sigma 0},~f_{\sigma 1},\cdots,~f_{\sigma 8})$; $\bm{S}$ is a non-negative diagonal matrix
\begin{equation}\label{S}
\bm{S}=diag(\tau_{\rho},~\tau_{e},~\tau_{\varepsilon},~\tau_{d},~\tau_{q},~\tau_{d},~\tau_{q},~\tau_{s},~\tau_{s})^{-1},
\end{equation}
in which all the relaxation times are associated with the moments: $\tau_{\rho}$ is related to the relax density, $\tau_{e}$ corresponds to the total energy, $\tau_{\varepsilon}$ is relevant to the energy square, $\tau_{d}$ associates with the momentum components, $\tau_{q}$ is related to the heat flux, and $\tau_{s}$ depends on the stress tensor. The discrete velocities $\bm{c}_{i}$ in Eq. \eqref{MRT} are given by $\bm{c}_{0}=(0,0),~\bm{c}_{1}=-\bm{c}_{3}=\bm{c}(0,1),~\bm{c}_{2}=-\bm{c}_{4}=\bm{c}(1,0),~\bm{c}_{5}=-\bm{c}_{7}=\bm{c}(1,1)$ and $\bm{c}_{6}=-\bm{c}_{8}=\bm{c}(-1,1)$. Here, the lattice speed $c$ is defined as $c=\delta_x/\delta_t$ with the lattice spacing $\delta_x$ and time step $\delta_t$, and is taken to be the velocity unit, i.e., $c=1$.

As the relaxation time parameters in the matrix $\bm{S}$ equal to the same value, the MRT model reduces to the Bhatnagar-Gross-Krook (BGK) model. The local equilibrium distribution function in either the BGK or the MRT model in Eq.~\eqref{MR} is given by:
\begin{equation}\label{feq}
f_{\sigma i}^{eq}=\omega_{i}\rho_{\sigma}\left[\alpha_{\sigma_{i}}+\frac{\bm{c_{i}}\cdot\bm{u}}{c^{2}_{s}}
+\frac{(\bm{c_{i}}\cdot\bm{u})^{2}}{2c^{4}_{s}}-\frac{\bm{u}^{2}}{2c^{2}_{s}}\right],
\end{equation}
where $\omega_{0}=4/9$, $\omega_{i}=1/9 (i=1-4) $, and $\omega_{i}=1/36 (i=5-8)$; $\rho_{\sigma}$ is the species density, $\bm{u}$ is the velocity of the gas mixture, $\alpha_{\sigma i}= s_{\sigma}=m_{min}/m_{\sigma}$ for $i\neq 0$ and $\alpha_{\sigma 0}= (9-5s_{\sigma})/4$ is a parameter dependent on the molecular mass $m_{\sigma}$ and the velocity $\bm{c}_i$, $m_{min}=min(m_a, m_b)$; $c_{s}=k_B T/m_{min}$ is a model-dependent parameter and equals to $c_s=c/\sqrt{3}$ here. The mass density $\rho$ and velocity $\bm{u}$ of the mixture, and the density $\rho_\sigma$ and velocity $\bm{u}_\sigma$ of the species are respectively defined as
\begin{eqnarray}\label{rhou}
 \rho=\sum_{\sigma}\sum_{i} f_{\sigma i},~~~\rho \bm{u}=\sum_{\sigma}\sum_{i} \bm{c}_{i}f_{\sigma i}, \nonumber \\
\rho_{\sigma}=\sum_{i} f_{\sigma i},~~~\rho_{\sigma} \bm{u}_{\sigma}=\frac{2\tau_{d}-1}{2\tau_{d}}\sum_{i} \bm{c}_{i}f_{\sigma i}+\frac{\rho_{\sigma}\bm{u}}{2\tau_{d}},
\end{eqnarray}
where $\tau_{d}$ is the relaxation time relating to the diffusion in mixtures \cite{Guo2009}.

For simulations of microscale gaseous flows, the relaxation times are essential for the adopted MRT-LBE model, and the Knudsen effect should be taken into account \cite{Guo2006,Guo2008}. In Ref. \cite{Guo2009}, Guo et al. generalized the MRT-LBE model for continuum binary mixtures \cite{ASina08} to microscale binary mixtures, and provided the relationship between the relaxation time and the Knudsen number. However, the range of Knudsen number is limited in this model. To extend the range of Knudsen number for microgas flows, the effective mean-free-path (MFP) scheme \cite{Guo2006,Guo2008}, which includes the termination effect from walls on the flight paths of gas molecules, would be incorporated to the LBE method. In this work, we will generalize the MFP scheme to the MRT-LBE model for micro-gaseous flows of binary mixtures so that the simulations can be reached at a wide range of Knudsen number. We would like to point out that such work has not been reported in the literature. In what follows, the effective relaxation times $\tau_s$ and $\tau_d$ will be given.

In Ref. \cite{Guo2009}, the relaxation time $\tau_s$ relates to the mean-free path $\lambda$ of binary mixture as
\begin{equation}\label{EQFrb}
  \lambda=\sqrt{\frac{3 \pi m_x}{2 m_{min}}}c_s^2\left(\tau_s-\frac{1}{2}\right)\delta_t,
\end{equation}
where $m_{x}$ is the molecular mass of gas mixture and defined as $m_x=\rho/n=x_am_a+x_bm_b$, and $x_\sigma=n_\sigma/n$ where $n=n_a+n_b$ and $n_\sigma$ are respectively the number density of the species and mixture.
Similarly, the relaxation time $\tau_d$ is related with the mean-free path of the species ( cf Ref. \cite{Guo2009} and reference therein ),
\begin{equation}\label{EQFrsp}
\left(\tau_d-\frac{1}{2}\right)\delta_t=\frac{3}{2}\sqrt{\frac{3m_0m_am_b}{m_{min}m_x^2}}\left[\frac{1}{\sqrt{x_a\lambda_a}}
  +\frac{1}{\sqrt{x_b\lambda_b}}\right]^{-2}
\end{equation}
where $m_0=m_a+m_b$, and $\lambda_\sigma$ is the mean-free path of specie $\sigma (a~and~b)$. From the kinetic theory, the mean-free path of the single-species $\sigma$ gas and the binary mixture can be determined by the dynamic viscosity \cite{Cercignani90,Yalamov94,Sharipov02}
\begin{equation}
     \lambda_\sigma=\frac{\mu_\sigma}{p_\sigma}\sqrt{\frac{\pi k_B T}{2m_\sigma}},\qquad
     \lambda=\frac{\mu}{p}\sqrt{\frac{\pi k_B T}{2m_x}},
\end{equation}
where $\mu_\sigma$  and $\mu$ denote the dynamic viscosity of the $\sigma$-species gas and the binary mixture, respectively.

For microgas flows with high Knudsen number, the flight path of some molecules will be cut off by the wall. Thus, the mean free path of gas molecules should be significantly influenced. The basic idea of the MFP scheme is to consider the effect of wall confinement on the mean free path through a correction function. With this point, the mean-free path in the left hand side of Eq. \eqref{EQFrb} should be replaced by the effective mean free path, which is expressed as
\begin{equation}
   \lambda_e=\lambda\Psi(y)=\lambda\frac{1}{2}\left(\psi\left(\frac{y}{\lambda}\right)+\psi\left(\frac{H-y}{\lambda}\right)\right),
\end{equation}
where $\lambda$ is taken as the mean free path for an unbounded system, $y$~($0\leq y \leq H$) is the distance from the particle to the nearest boundary, and the function $\psi$ is defined by
\begin{eqnarray}\label{psi}
\psi(\alpha)=1+(\alpha-1)e^{-\alpha}-\alpha^{2}E_i(\alpha),\\
 E_i(x)=\int_{1}^{\infty} t^{-1}e^{-x t}dt.
\end{eqnarray}
Thus, via Eq. \eqref{EQFrb} with the effect mean free path, the effective relaxation time $\tau_{s}$ in the present work is determined by
\begin{equation}\label{taos}
\tau_{s}=\frac{1}{2}+\frac{\lambda\Psi(y)}{\delta_t}\sqrt{\frac{6m_{min}}{\pi m_{x}}}.
\end{equation}
The effective relaxation time $\tau_{d}$ is similarly determined from Eq. \eqref{EQFrsp} using the effective mean free path $\lambda_{ae}$ and $\lambda_{be}$
\begin{equation}\label{EQFrsp}
   \tau_d =\frac{1}{2}+\frac{3}{2\delta_t}\sqrt{\frac{3m_0m_am_b}{m_{min}m_x^2}}\left[\frac{1}{\sqrt{x_a\lambda_{ae}}}
  +\frac{1}{\sqrt{x_b\lambda_{be}}}\right]^{-2},
\end{equation}
where $\lambda_{\sigma e}=\lambda_\sigma\Psi(y)=\lambda_\sigma\left(\psi\left(\frac{y}{\lambda_\sigma}\right)+\psi\left(\frac{H-y}{\lambda_\sigma}\right)\right)/2$. As pointed out in Ref. \cite{Guo2009}, the relaxation time $\tau_e$ is determined according to the bulk viscosity, and the other relaxation parameters can be adjusted with much freedom for better numerical stability. It should be seen that the values of $\tau_s$ and $\tau_d$ can be determined turning to the Knudsen number of the mixture and/or species since $Kn_i=\lambda_i/H$.
\subsection{Boundary conditions}
\label{boundary}
Boundary conditions play an important role in the LBE method for microscale gas flows~\cite{ Tang2004, Succi2005, Guo2006, Guo2008, Li2009}. There are several schemes to realize the slip boundary conditions in the case of single-species gas flows, such as the bounce-back--specular-reflection (BSR) scheme \cite{Succi2005}, the discrete Maxwell's diffuse--reflection (DMDR) scheme, and the diffuse-bounce-back (DBB) scheme~\cite{Guo2011}. In Ref.~\cite{Guo2009}, the BSR scheme was extended to binary gas flows. However, the BSR and DMDR scheme are non-local due to the inclusion of the specular-reflection, and this may bring difficulties in simulating micro-flows with complex geometries. Note that the information of one point is only needed to determine the unknown distribution function at the same point in the DBB scheme. In this work, we will propose a DBB scheme for the binary mixture model
\begin{figure}
\centering
\includegraphics[scale=0.95]{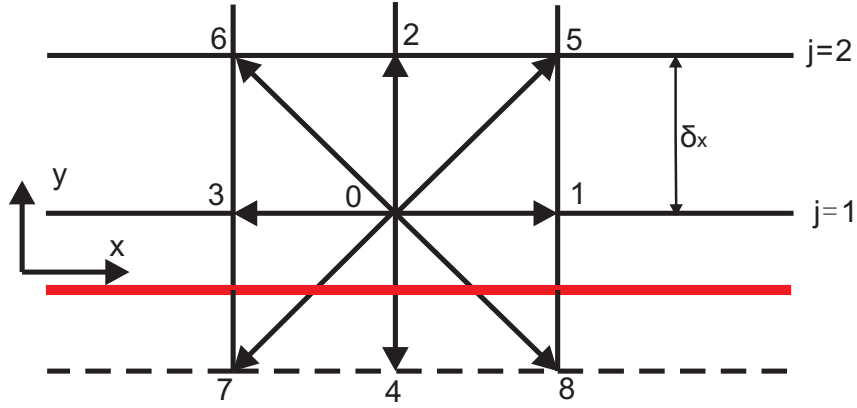}
\caption{Lattice and boundary arrangement in the D2Q9 model. The bottom wall is located at $j=1/2$.}
\label{Fig:boundary}
\end{figure}
At the bottom boundary as shown in Fig.~\ref{Fig:boundary}, the unknown distribution functions are set as
\begin{eqnarray}\label{bot}
\begin{cases}
  f_{\sigma,2}=r_{\sigma}f_{2}^{eq}(\bm{u}_w)+(1-r_{\sigma})f^{'}_{\sigma,4}~,\\
  f_{\sigma,5}=r_{\sigma}f_{5}^{eq}(\bm{u}_w)+(1-r_{\sigma})f^{'}_{\sigma,7}~,\\
  f_{\sigma,6}=r_{\sigma}f_{6}^{eq}(\bm{u}_w)+(1-r_{\sigma})f^{'}_{\sigma,8}~.
\end{cases}
\end{eqnarray}
and at the top plate,
\begin{eqnarray}\label{bot2}
\begin{cases}
f_{\sigma,4}=r_{\sigma}f_{\sigma,4}^{eq}(\bm{u}_w)+(1-r_{\sigma})f^{'}_{\sigma,2}~,\\
f_{\sigma,7}=r_{\sigma}f_{\sigma,7}^{eq}(\bm{u}_w)+(1-r_{\sigma})f^{'}_{\sigma,5}~,\\
f_{\sigma,8}=r_{\sigma}f_{\sigma,8}^{eq}(\bm{u}_w)+(1-r_{\sigma})f^{'}_{\sigma,6}~.
\end{cases}
\end{eqnarray}
where $r_{\sigma}$ is the parameter representing the equilibrium part, $\bm{u}_w$ is the velocity of the mixture on the boundary, and $f^{'}_{i}$ is the post-collision distribution function. In order to realize the slip boundary condition exactly, $r_{\sigma}$ and $\tau_{q}$ should be carefully chosen. Following the derivations of Refs. \cite{Guo2009,Guo2008}, these two parameters in the present DBB scheme are taken as
\begin{equation}\label{R}
  r_{\sigma}=2-2\left[1+\frac{c_{m}}{3}\sqrt{\frac{\pi m_{x}}{2k_B T}}\right]^{-1}=2-2\left[1+c_{m}\sqrt{\frac{\pi m_{x}}{6m_{r}}}\right]^{-1},
\end{equation}
and
\begin{equation}\label{taoq}
\tau_{q}=0.5+\frac{3+24\varsigma^{2}\widetilde{\tau}_{s}^{2}(0)A_{2}}{16\widetilde{\tau}_{s}(0)}+\frac{\tau_{s}(0)\delta_{x}[12+30\widetilde{\tau}_{s}(0)\varsigma
A_{1}]}{16\widetilde{\tau}_{s}^{2}(0)},
\end{equation}
where \textbf{$m_{r}=m_{\sigma}/m_{x}$}, $\varsigma=\sqrt{\pi/6}$, $\widetilde{\tau_{s}}(0)=\tau_{s}(0)-0.5$, $\delta_{x}=H/N$ with $N$ being the lattice number in the characteristic length $H$, and $\tau'_{s}(0)=\partial y\tilde{\tau_{s}}(0)$ in which $0$ refers to the position at the boundary. $A_{1}$ and $A_{2}$ are given by
\begin{equation}\label{A12}
A_{1}=\frac{2-\alpha}{\alpha}(1-0.1817\alpha),~~A_{2}=\frac{1}{\pi}+\frac{1}{2}A_{1}^{2},
\end{equation}
where  $0<\alpha\leq 1$ is the accommodation coefficient. The parameter $c_{m}$ is the velocity slip coefficient (VSC), which is gained from the linearized Boltzmann equation \cite{Ivchenko1997}
\begin{equation}\label{cm}
c_{m}=\frac{PM^{1/2}_{s}}{\mu}\frac{5\pi}{8}\sum\limits_{\sigma}\left[(2-\alpha)x_{\sigma}b_{\sigma}\left(K_{1}+\frac{4b_{\sigma}}{\pi M_s^{1/2}}K_{2}\right)\right],
\end{equation}
where $M_s=m_{x}/m_{0}$ and $b_{\sigma}$ is given by:
\begin{equation}\label{bm}
b_{\sigma}=\frac{x_{\sigma}R_{\sigma}+x_{\varsigma}T_{\varsigma}}{P(x_{a}^{2}R_{a}/\mu_{a}+x_{b}^{2}R_{b}/\mu_{b}+x_{b}x_{a}R_{ab})},
\end{equation}
with $K_1$ and $K_2$ being defined by
\begin{equation}\label{K}
K_{1}=\frac{\sum\limits_{\sigma}(2-\sigma)x_{\sigma}b_{\sigma}}{\sum\limits_{\sigma}\alpha_{\sigma}x_{\sigma}M_s^{1/2}}K_{2},\quad K_{2}=\frac{p}{4\mu}.
\end{equation}

For the pressure boundary conditions at the inlet and outlet, we will employ an extrapolation-correction technique which has been used for single-specie gas flows~\cite{Xu2013,Guo2006}. Concretely speaking, at the inlet ($x=0$), the unknown distribution functions and density are obtained by extrapolation from the inner nodes
\begin{equation}\label{rho1}
f_{\sigma i}(0,j)=2f_{\sigma i}(1,j)-f_{\sigma i}(2,j),~~~\rho^{'}_{\sigma}(0,j)=2\rho_{\sigma}(1,j)-\rho_{\sigma}(2,j),
\end{equation}
where 0, 1, and 2 denote the nodes at the inlet, first layer, and second layer, respectively. The density is then corrected so that the average density $\rho_{in}$ matches that given by the pressure boundary condition
\begin{equation}\label{rho2}
\rho_{\sigma}(0,j)=\rho_{\sigma}^{'}(0,j)\frac{N\rho_{\sigma, in}}{\sum\limits_{j}{\rho^{'}_{\sigma}}(0,j)},
\end{equation}
where $\rho_{\sigma, in}=p_{\sigma,in}/c_s^2$. The distribution functions and densities at the outlet can be obtained in a similar way.

\subsection{Method validation}
\begin{figure}
\centering
\includegraphics[width=0.7\textwidth]{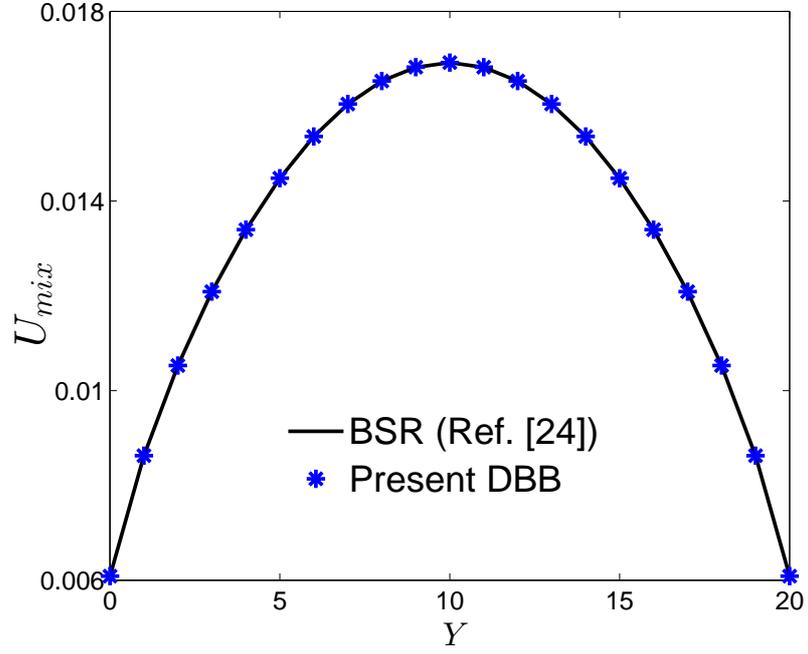}
\caption{Mixture velocity profile of the gaseous flows of He--Ar mixtures for Case 2 at $\textrm{Kn}_{out}$=0.3 and $\theta=3$.}
\label{ub}
\end{figure}

\begin{figure}
\centering
\includegraphics[width=0.66\textwidth]{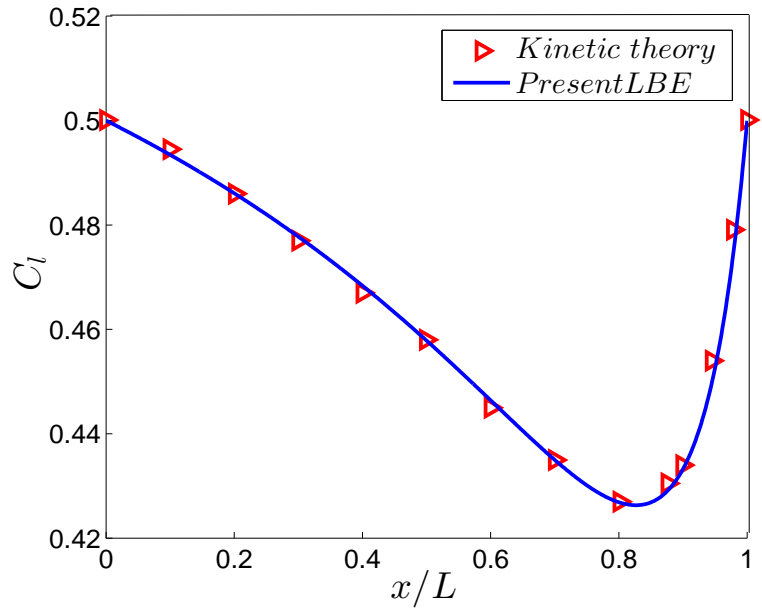}
\caption{The mole fraction of the light species $C_{l}$ along the channel for Case 2 at $\theta=2.0$ and $\textrm{Kn}_{out}=0.3$. Symbols: the results obtained by the McCormack kinetic method \cite{Szalmas}; Solid lines: the present MRT-LBE results.}
\label{eop}
\end{figure}

In order to verify the proposed DBB scheme, the MRT-LBE together with the DBB and BSR boundary conditions are tested by simulations of Case 2 with $\textrm{Kn}_{out}=0.3$ and $\theta=3$. In all of the following simulations, the relaxation times are set as follows: $\tau_{s}$ is determined by Eq.~\eqref{taos}, $\tau_q$ is given by Eq.~\eqref{taoq}, $\tau_{d}$ is obtained from Eq. \eqref{EQFrsp}, $\tau_{\rho}=1$ for conserved variables, and the remains are given by $\tau_{e}=1.1$ and $\tau_{\varepsilon}=1.2$. The gas mixture velocity is shown in Fig.~\ref{ub} along with the results from the BSR scheme in Ref.~\cite{Guo2009}. It can be seen that the results of the DBB scheme is in good agreement with that of BSR scheme for this problem.
Furthermore, the mole fraction of the light species along the channel is computed for Case 2 with $\textrm{Kn}_{out}=0.3$ and $\theta=2.0$. The results are shown in Fig. \ref{eop} where the results obtained by the McCormack kinetic method \cite{Szalmas} are included for comparison.
As can be seen, the numerical outcome from the present MRT-LBE model coincides excellently with that from the McCormack kinetic method. All of these favorable comparisons lend confidence in the accuracy of the MRT-LBE model together with the proposed DBB boundary conditions.

\section{Numerical results and discussions}
\label{annlyze}
In this section, the present MRT-LBE model together with the proposed DBB boundary condition are applied to the He/Ar and Ne/Ar gas mixtures at the cases listed in Table \ref{C3table1}. A parametric study will be performed to investigate the gas separation phenomenon affected by the pressure ratio, the Knudsen number and molecular mass and mole fraction. As an indicative parameter for the gas separation, the mole fraction in the binary mixture will be mainly considered in the subsequent analysis.

\subsection{Effect of pressure ratio}\label{pressure}
We first focus on the effect of pressure ratio on the separation process. As can be observed from Fig. \ref{eop}, the mole fraction of the light species $C_{l}$ is non-uniform along the channel, which decreases from the inlet first and then increases to the value at the outlet. This means that a local minimum exists in the distribution of $C_l$ at certain point in the channel. Consequently, via this minimum value and Eq. \eqref{CL}, the parameters $CL$ can be obtained to reflect the degree of separation.
\begin{figure}
\begin{tabular}{ccc}
\includegraphics[width=0.55\textwidth,height=0.325\textheight]{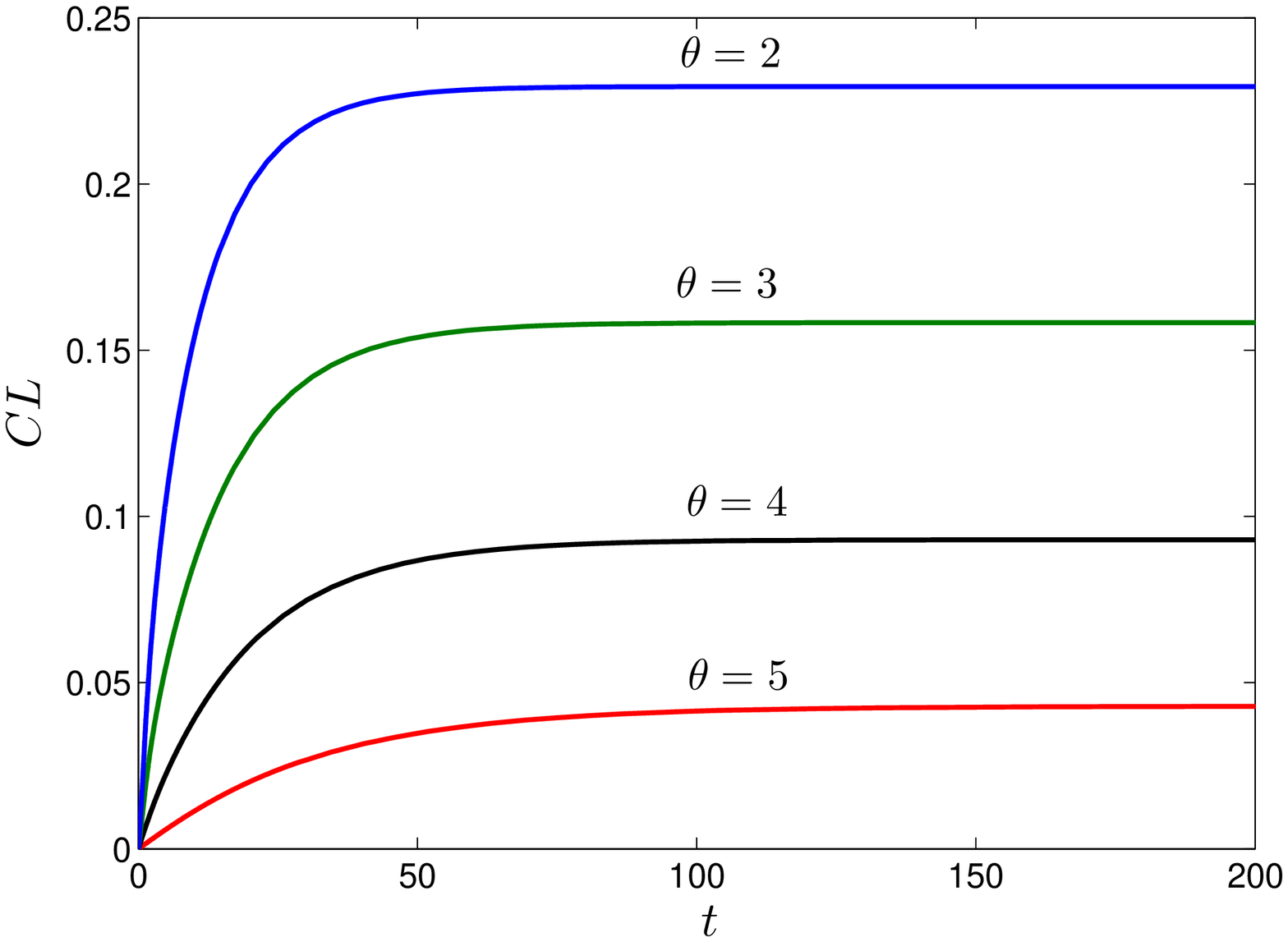}&
\includegraphics[width=0.55\textwidth,height=0.325\textheight]{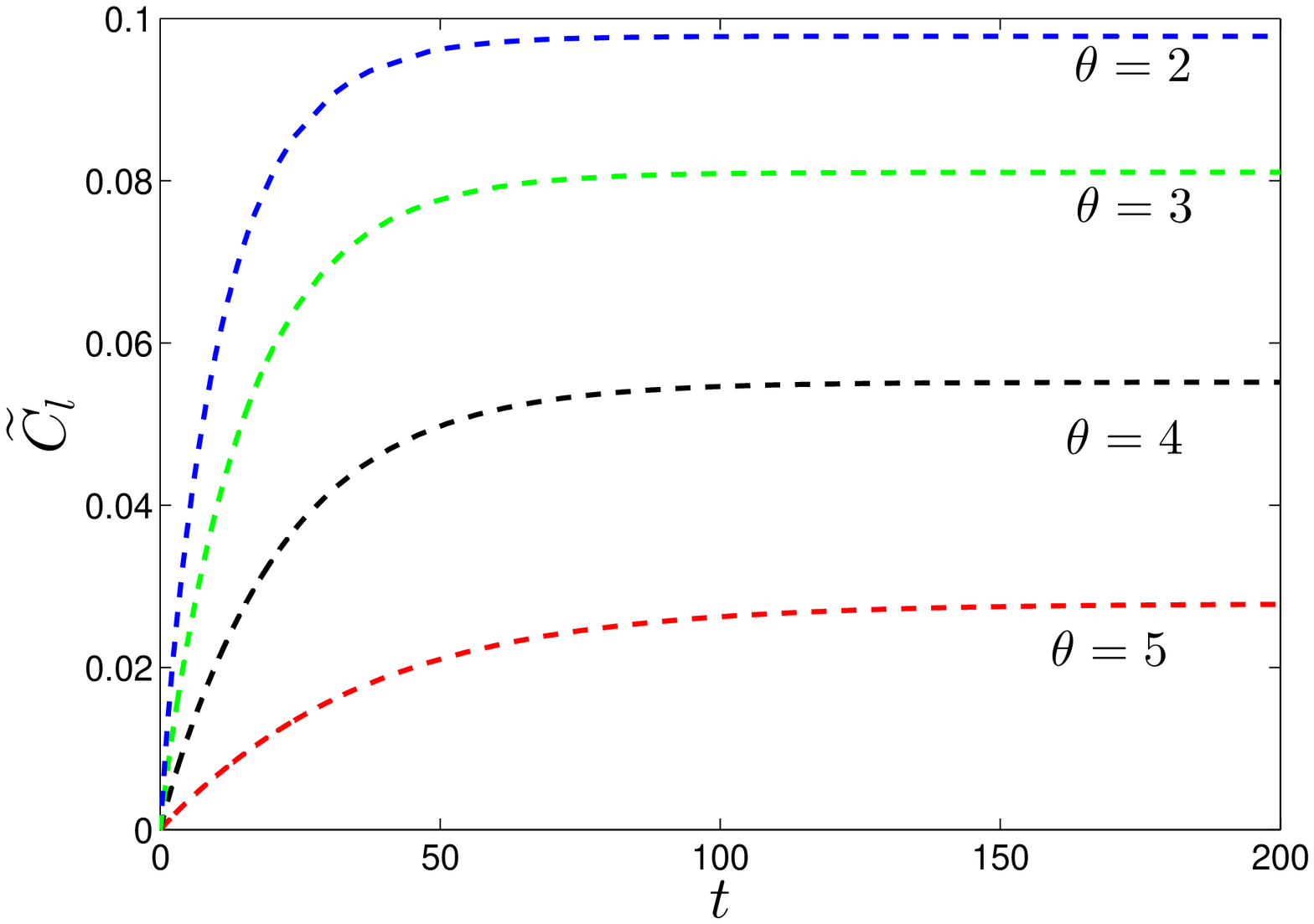}\\
(a)&(b)\\
\end{tabular}
\caption{Time history of (a) $CL$ and (b) $\widetilde{C_{l}}$ with different $\theta$ for Case 2 at $Kn_{out}=0.3$.}
\label{eop2}
\end{figure}
A number of simulations are carried out for different pressure ratios, and it is found that the numerical results in the He--Ar and Ne--Ar gas mixtures display a similar tendency on $CL$ and $\widetilde{C_{l}}$. Thus, the results of the He--Ar mixture are only shown here. In Fig. \ref{eop2}, the time history of $CL$ and $\widetilde{C_{l}}$ are presented at different $\theta$ for Case 2 with $Kn_{out}=0.3$ by plotting $CL$ and $\widetilde{C_{l}}$ versus time $t$ for different $\theta$, where the dimensionless time $t$ is defined as $t=n\delta t\times u_{c}/H$ in which $n$ is the computational time step, and $u_{c}$ is the velocity of the gas mixture obtained from the results of Case 2 with $\textrm{Kn}_{out}=0.3$ and $\theta=2.0$. As displayed in Fig. \ref{eop2} (a), the separation process can be clearly described by the change of $CL$ with time. In the initial time stage, the value of $CL$ at each $\theta$ increases quickly, which implies the commencement of gas separation. Subsequent to this stage, $CL$ remains with an unchanged value, and this indicates the termination of the separation process and the steady state of binary mixtures. Similar results can also be reflected from the time-evolution of $\widetilde{C_{l}}$ as shown in Fig. \ref{eop2} $(\textit{b})$.

From the curve plots shown in Fig. \ref{eop2}, we can also make the following interesting observations. That is, the smaller the pressure ratio $\theta$ is, the quicker the value of $CL$ (and $\widetilde{C_l}$) increases initially, and the larger the value of $CL$ is in the whole gas separation process. These mean that as the pressure ratio $\theta$ decreases, the rate of gas separation ($V_{s}$), which is represented by the curve gradient in the figure, increases, and the quality of gas separation increases. This tendency can be explained by the fact that with the decrease in $\theta$ for a given $\textrm{Kn}_{out}$, the average pressure of the mixture will decrease such that the mean Knudsen number increases, and the momentum exchange between the light and heavy molecules is reduced. Thus, the difference in the molecular velocities of different species will be enhanced and more obvious gas separation tends to occur. Therefore, the increase in the pressure ratio has a negative effect on the gas separation in the binary mixtures of gaseous flows.

\subsection{Effect of Knudsen number}\label{kn}
\begin{figure}
\begin{tabular}{ccc}
\includegraphics[width=0.5\textwidth,height=0.315\textheight]{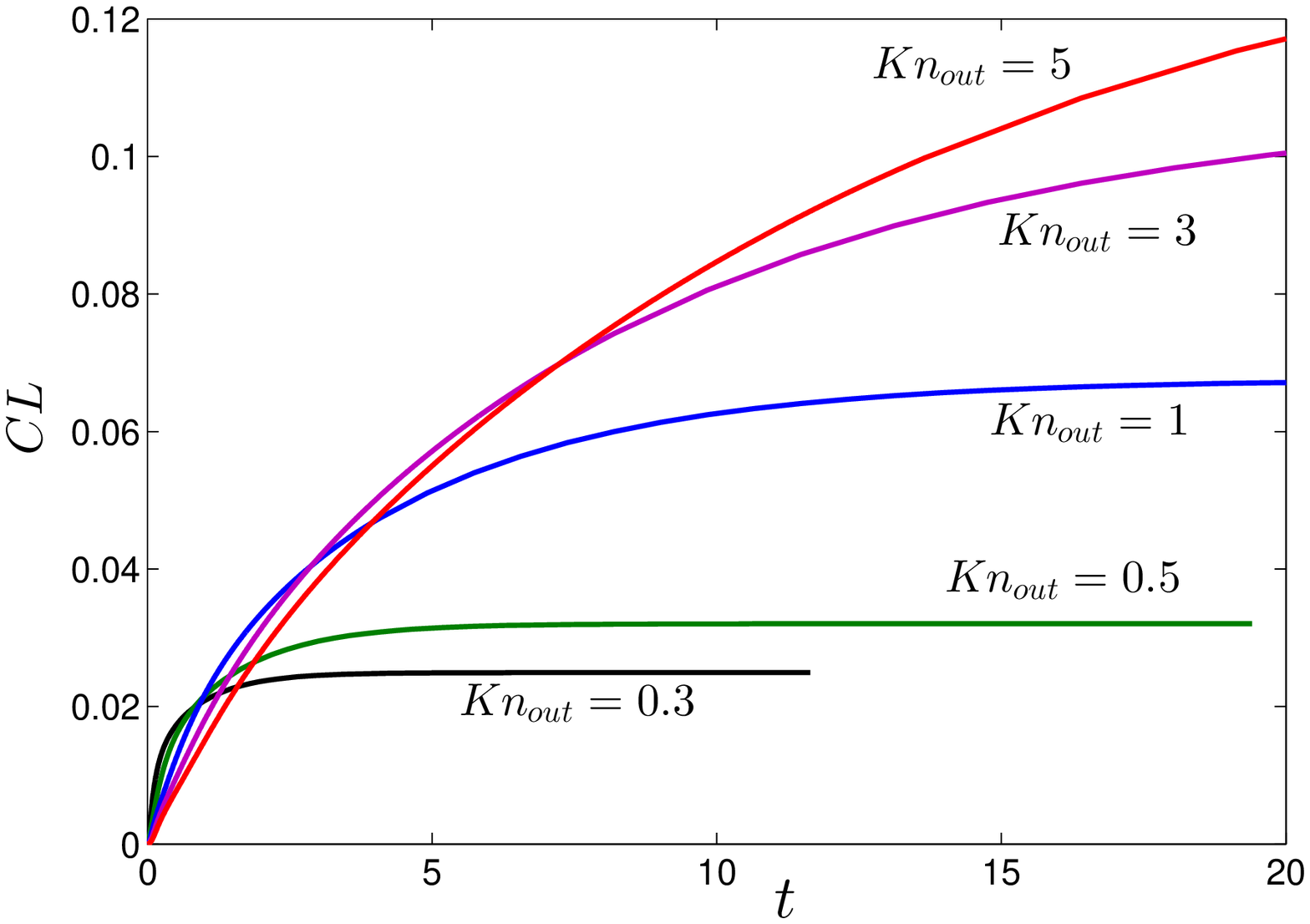}&
\includegraphics[width=0.5\textwidth,height=0.315\textheight]{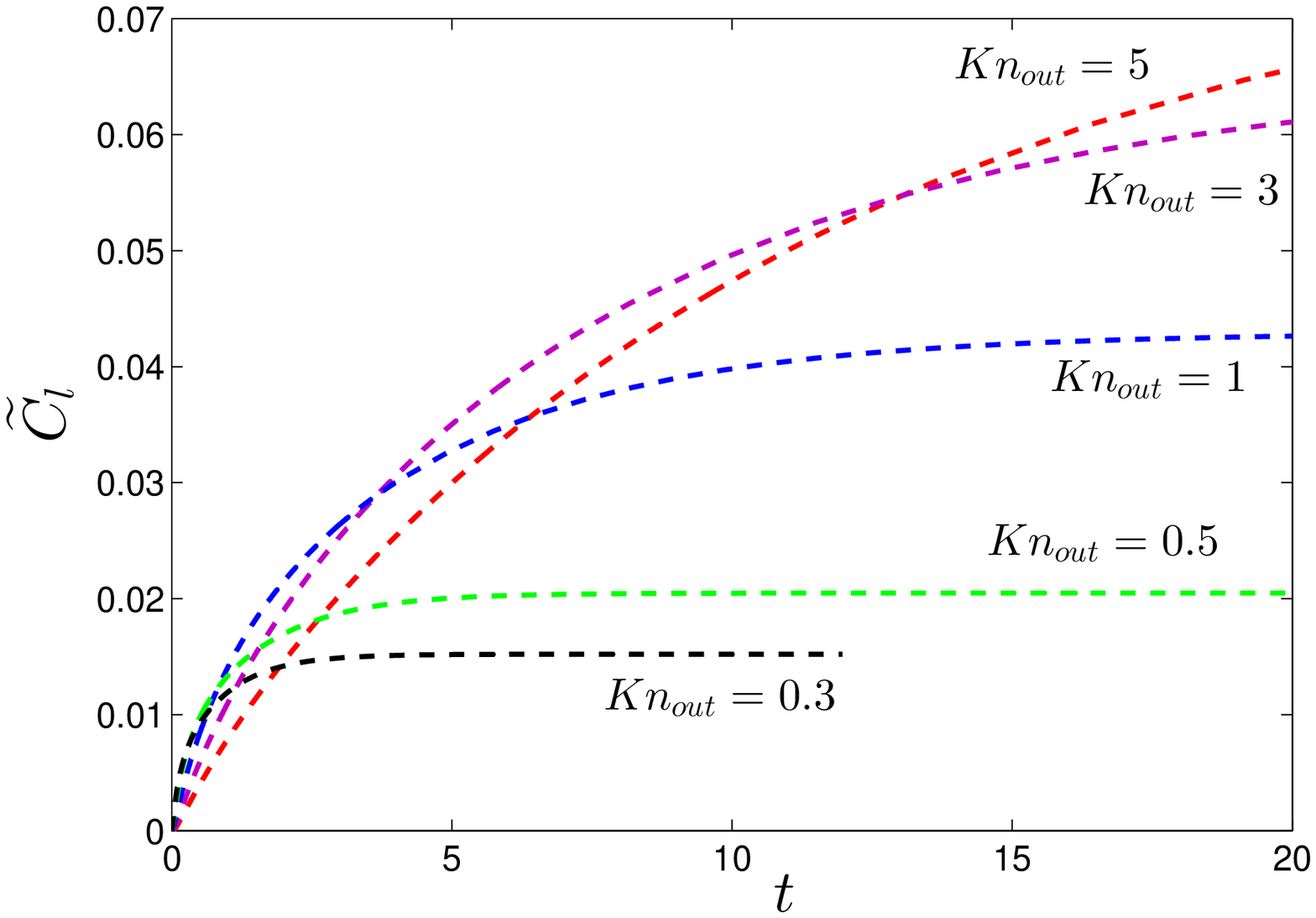}
\end{tabular}
\caption{Time evolution of $CL$ and $\widetilde{C_{l}}$ with different Knudsen numbers for Case 2 at $\theta=2.0$.}
\label{eok}
\end{figure}

\begin{figure}
\centering
\includegraphics[width=0.65\textwidth]{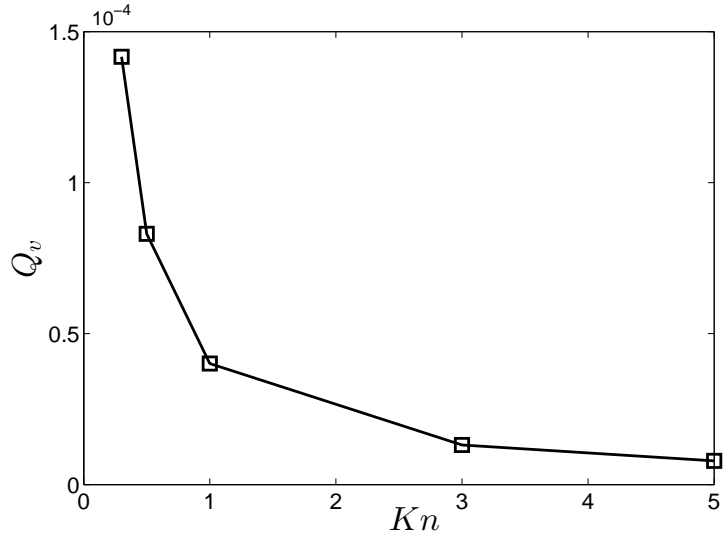}
\caption{Volume flow $Q_v$ of the binary mixtures against $\textrm{Kn}_{out}$ at $\theta=2.0$ for Case 2.}
\label{vflux}
\end{figure}

The rarefaction effect also plays a significant role in the microflows \cite{Karniadakis}, which can be modeled by different Knudsen numbers. Next, the effect of $Kn_{out}$ is investigated on the gas separation. Simulations with different $Kn_{out}$ at $\theta=2.0$ are performed still for Case 2. The results of the separation process are shown in Fig. \ref{eok}.
As can be observed, the terminal concentrations $CL$ and $\widetilde{C_l}$ increase as $\textrm{Kn}_{out}$ increases from $0.3$ to $5$. This is because a larger Knudsen number makes the gas mixture more rarefied for a given pressure ratio, and the momentum transfer between the light and heavy molecules becomes smaller. This will increase the velocity difference between the two gas molecules, and hence enhance the gas separation. In addition, one can find that the separation rate $V_{s}$ decreases slightly with increasing $\textrm{Kn}_{out}$. It is interesting to note that this tendency is opposite to that of increasing the Knudsen number induced by decreasing the pressure ratio. This is due to the fact that the increase in $\textrm{Kn}_{out}$ leads to the reduction of the flow velocity of binary mixtures, which yields a decrease in the time needed for completing the gas separation. To affirm this point, we compute the volume flow $Q_v$ of the binary mixtures and show the results in Fig. \ref{vflux}.
Clearly, the volume flow $Q_v$ decrease with the increase in the Knudsen number $\textrm{Kn}_{out}$.

\subsection{Effect of molecular mass and mole fraction}\label{mass}
The mole fraction and molecular mass are two particular parameters consisting of the gas mixtures, and they shall have a significant effect on the gas separation. \begin{figure}
\begin{tabular}{ccc}
\includegraphics[width=0.5\textwidth,height=0.31\textheight]{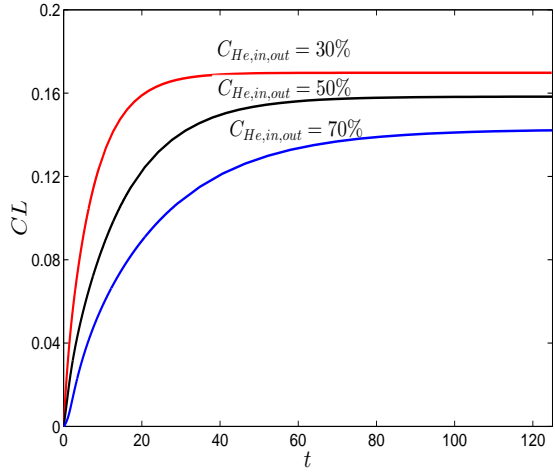}&
\includegraphics[width=0.5\textwidth,height=0.31\textheight]{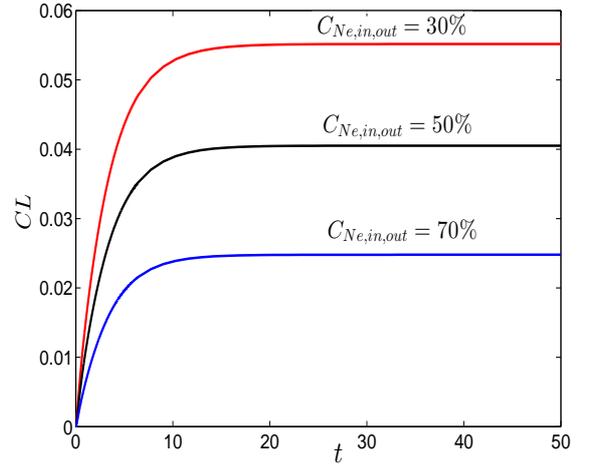}\\
(a)&(b)\\
\includegraphics[width=0.5\textwidth,height=0.31\textheight]{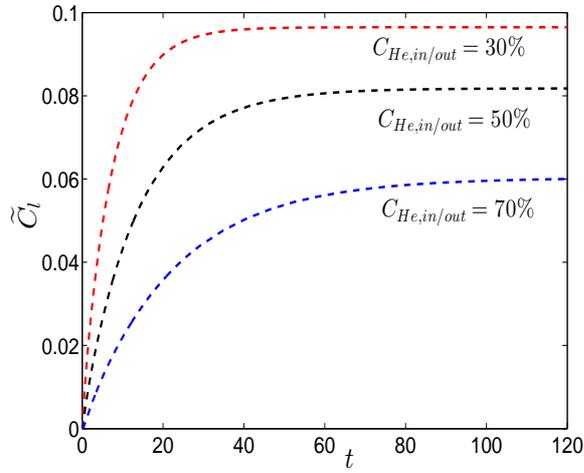}&
\includegraphics[width=0.5\textwidth,height=0.31\textheight]{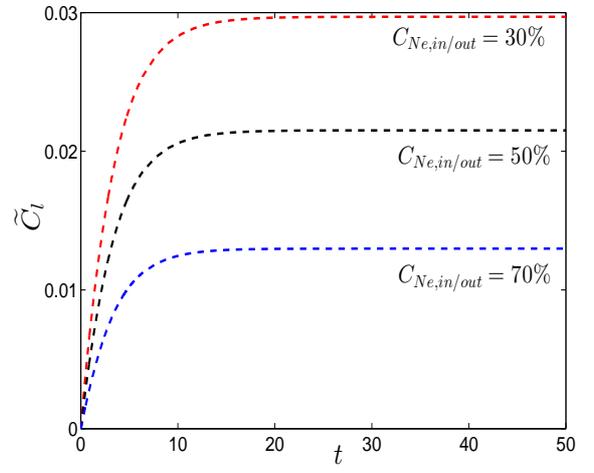}\\
(c)&(d)
\end{tabular}
\caption{Time history of $CL$ and $\widetilde{C_l}$ with different values of $C_{l, in, out}$ ($C_{l, in}=C_{l, out}$) at $\textrm{Kn}_{out}=0.3$ and $\theta=2.0$ for (a) and (c): the He--Ar mixture; (b) and (d): the Ne--Ar mixture.}
\label{eom}
\end{figure}
In Fig. \ref{eom}, the changes of $CL$ and $\widetilde{C_l}$ with time are plotted at $\textrm{Kn}_{out}=0.3$ and $\theta=2.0$ for Cases 1--3 and 6--8. In each case, the concentrations of the He--Ar or Ne--Ar mixture at the inlet and outlet are the same. As shown in the figure, $CL$ and $\widetilde{C_l}$ decrease in the concentration of $C_{l, in}$ (or $C_{l, out}$) for the two binary mixtures systems, which indicates the increase of the separation degree. It is also found that the separation rate $V_s$ has a similar tendency with the values of $C_{l, in}$. This may be attributed to the fact that a decrease in $C_{l, in}$ can lead to a decrease in the overall mole mass of the gas mixture under the same $\theta$ and $\textrm{Kn}_{out}$, which boosts the refraction of the gaseous mixtures, and thereby yields an increase in the separation rate.

The second observation in Fig. \ref{eom} comes from the comparison of the results between the He--Ar mixture and the Ne--Ar mixture, which shows a clear difference in $CL$. It is found that the terminal $CL$ in the He--Ar mixture is much larger than that in the Ne--Ar mixture (see Fig.~\ref{eom} (a) and (c)). This indicates us that the gas separation is more thorough for the He--Ar mixture compared with the Ne--Ar mixture. The reason for this difference is that the larger molecular mass ratio brings about the larger velocity difference between the species in the He--Ar system, and thus enhances the quality of gas separation. On the other hand, it can be seen from Figs. \ref{eom} (a) and (b) that $CL$ in the Ne--Ar mixture increases to the state-state value faster than that in the He--Ar mixture. This further demonstrates the better quality of gas separation in the He--Ar mixture than the Ne--Ar mixture.

\begin{figure}
\begin{tabular}{ccc}
\includegraphics[width=0.5\textwidth,height=0.325\textheight]{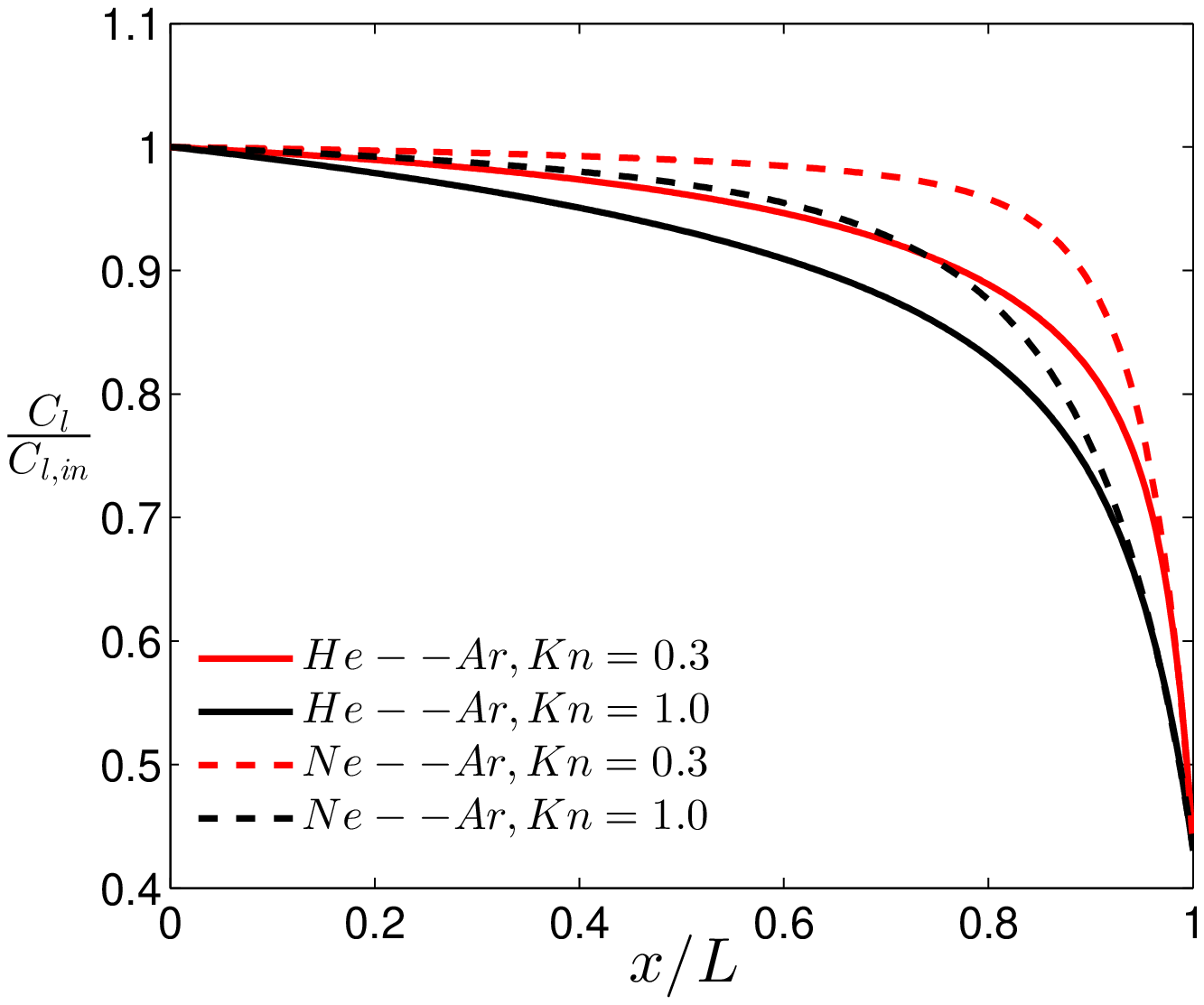}&
\includegraphics[width=0.5\textwidth,height=0.325\textheight]{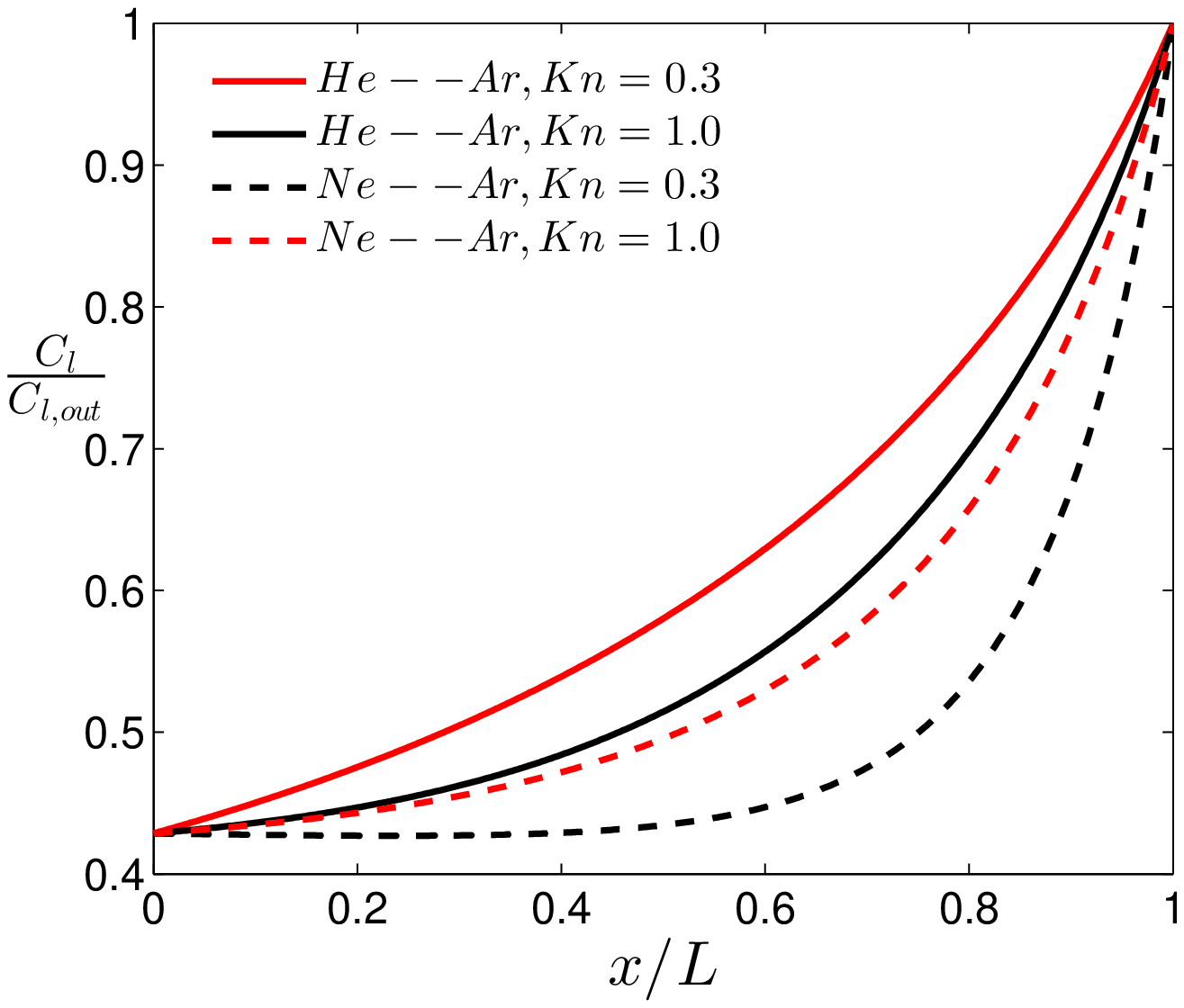}\\
(a)&(b)\\
\end{tabular}
\caption{The dimensionless mole fraction of $C_{l}$ along the channel with different Knudsen numbers and mole fractions at the inlet and outlet. (a): $C_{l, in}$=0.7, $C_{l, out}$=0.3. (b): $C_{l, in}$=0.3, $C_{l, out}$=0.7.}
\label{eom2}
\end{figure}

Then, to highlight the effect of mole fraction at the inlet and outlet, the computational parameters of Cases 4, 5, 9 and 10 listed in Table \ref{C3table1} are employed. Fig. \ref{eom2} presents the curve profile of dimensionless mole fraction $C_{l}$ along the channel.
We can see from both subfigures that the mole fraction distribution along the channel is nonlinear, which is similar to the results found by Kalempaa and Sharipov \cite{Kalempaa}. Through comparisons of mole fraction at different Knudsen numbers in one binary mixture as well as at the same Knudsen number but in different binary mixtures, it can be found that the increase of Knudsen number and molecular mass ratio can lead to the gas separation phenomenon easier, which is consistent with the foregoing results. Particularly, when careful attention is paid to the mole fraction $C_l$ near the inlet in Fig. \ref{eom2} (a) and the outlet in Fig. \ref{eom2} (b), it is found that $C_l$ in these regions varies much faster than in other places of the channel, which can actually be deduced from the convexity (Fig. \ref{eom2} (a)) and concavity (Fig. \ref{eom2} (b)) properties of $C_l$ as well. This indicates the appearance of the separation phenomenon, and more importantly, this leads us to an interesting result that the gas separation takes place mainly in the region near the channel end with higher mole fraction. It is noteworthy that this finding has not been reported in previous studies.

\section{Conclusions} \label{results}
In this paper, the separation phenomenon of a binary gaseous mixture is investigated by the MRT-LBE method with a proposed kinetic boundary condition combining the bounce-back and the discrete Maxwell's diffuse schemes. The wall-confinement effects are considered in the effective relaxation times to simulate microscale flows in both slip and transition regimes. It is seen that the mole fraction of each component in the gas mixtures does not distribute uniformly along a microchannel. At an equal mole fraction given at the channel ends, there will be a minimum mole fraction for the lighter species in the channel. For the separation process, the effects of pressure ratio, rarefaction, mole fraction and molecular mass ratio are studied in detail. It is found that the degree and rate of gas separation are influenced by these parameters, and the main results are summarized as follows:

First, the separation process is fierce as the pressure ratio decreases. This is because a smaller pressure ratio enhances the rarefaction, which leads to a decrease in the momentum transfer between the light and heavy molecules and further increases the difference in species velocities.

Second, the separation phenomenon is enhanced when the Knudsen number becomes larger in that the resulted larger rarefaction effect increases the difference in species velocities. On the other hand, the separation rate decreases slightly with increasing the Knudsen number due to the reduced molecular velocity.

Finally, the separation process is more obvious for a mixture with large molecular mass ratio as expected, and the degree and rate of separation increase as the mole fraction of the light species decreases. Particularly and interestingly, the channel end with higher mole fraction of lighter species is the main region where the gas separation takes place.

The present work on the gas separation phenomenon is limited to the two dimension case, which is inadequate to reveal the fundamental knowledge. In the future, we will extend this work to three dimensional cases, and a detailed parametric study on the gas separation will be followed.

\section*{Acknowledgments}
This work is financially supported by the National Natural Science Foundation of China (Grant Nos. 51125024 and 51390494).


\begin{thebibliography}{99}
\bibitem{Karniadakis} G.E. Karniadakis, A. Beskok, Micro Flows: Fundamentals and Simulation, Springer, 2001.
\bibitem{Squires} T.M. Squires, S.R. Quake, Microfluidics: Fluid physics at the nanoliter scale, Rev. Mod. Phys. 77 (2005) 977-1026.
 \bibitem{Nie}X.B. Nie, G.D. Doolen, S.Y. Chen, Lattice-Boltzmann Simulations of Fluid Flows in MEMS, J. Stat. Phys. 107 (2002) 279-289.
\bibitem{Chapman1970} S. Chapman, T. G. Cowling, The Mathematical Theory of Non-Uniform Gases, 3rd ed. Cambridge University Press, Cambridge, England, 1970.
\bibitem{Zhang2011} J.F. Zhang, Lattice Boltzmann method for microfluidics: models and applications, Microfluid. Nanofluid. 10 (2005) 1-28.
\bibitem{Wu} P.Y. Wu, W.A. Little, Measurement of friction factors for the flow of gases in very fine channels used for microminiature Joule-Thomson refrigerators, Cryogenics. 23 (1983) 273-277.
\bibitem{Present} R.D. Present, A.J. Debethune, Separation of a Gas Mixture Flowing through a Long Tube at Low Pressure, Phys. Rev. 75 (1949) 1050-1057.
\bibitem{Szalmas} L. Szalmas, J. Pitakarnnop, S. Geoffroy, S. Colin, D. Valougeorgis, Comparative study between computational and experimental results for binary rarefied gas flows through long microchannels, Microfluid. Nanofluid. 9 (2010) 471-487.
\bibitem{Szalmas2} L. Szalmas, D. Valougeorgis, Rarefied gas flow of binary mixtures through long channels with triangular and trapezoidal cross sections, Microfluid. Nanofluid. 9 (2010) 1103-1114.
\bibitem{Myong} R.S. Myong, A generalized hydrodynamic computational model for rarefied and microscale diatomic gas flows, J. Comput. Phys. 195 (2004) 655-676.
\bibitem{Dodulad} O.I. Dodulad, I.D. Ivanova, Y.Y. Kloss, P.V. Shuvalov, F.G. Tchremissine, Study of gas separation in micro devices by solving the Boltzmann equation, 28th International Sympsium on Rarefied Gas Dynamics (2012) 816-823.
\bibitem{Sharipov05}F. Sharipov, D. Kalempa, Separation phenomena for gaseous mixture flowing through a long tube into vacuum, Phys. Fluids. 17 (2005) 127102.
\bibitem{Takata07}S. Takata, H. Sugimoto, S. Kosuge, Gas separation by means of the Knudsen compressor, Eur. J. Mech. B/Fluids. 26 (2007) 155-181.
\bibitem{Varoutis09}S. Varoutis, S. Naris, V. Hauer, C. Day, D. Valougeorgis, Experimental and computational investigation of gas flows through long channels of various cross sections in the whole range of the Knudsen number. J. Vac. Sci. Technol. A. 27(1) (2009) 89-100.
\bibitem{Vargo} S.E. Vargo, E.P. Muntz, G.R. Shiflett, W.C. Tang, Knudsen compressor as a micro- and macroscale vacuum pump without moving parts or fluids, J. Vac. Sci. Technol. A 17 (1999) 2308-2313.
\bibitem{McNamara} S. McNamara, Y.B. Gianchandani, A micromachined Knudsen pump for on-chip vacuum, 12th International Conference on Transducers, Solid-State Sensors, Actuators And Microsystems 2 (2003) 1919-1922.
\bibitem{Aoki} K. Aoki, P. Degond, L. Mieussens, Numerical simulations of rarefied gases in curved channels: thermal creep, circulating flow, and pumping effect, Commun. Comput. Phys. 6 (2009) 919-954.
\bibitem{Higashi64}K. Higashi, H. ITO, J. OISHI, Surface diffusion phenomena in gaseous diffusion, (II) Separation of Binary Gas-mixtures, J. Nucl. Sci. Tech. 1 (1964) 298-304.
\bibitem{Kalempaa} D. Kalempaa, F. Sharipov, Flows of rarefied gaseous mixtures with a low mole fraction. Separation phenomenon, Euro. J. Mech. B/Fluids 30 (2011) 466-473.
\bibitem{Szalmas10}L. Szalmas, D. Valougeorgis, Rarefied gas flow of binary mixtures through long channels with triangular and trapezoidal cross sections, Microfluid. Nanofluid. 9 (2010) 471-487.
\bibitem{Arcidiac07}S. Arcidiacono, I.V. Karlin, J. Mantzaras, C.E. Frouzakis, Lattice Boltzmann model for the simulation of multicomponent mixtures, Phys. Rev. E 76 (2007) 046703.
\bibitem{Joshi07}A.S. Joshi, A.A. Peracchio, K.N. Grew, W.K.S. Chiu, Lattice Boltzmann method for multi-component, non-continuum mass diffusion, J. Phys. D: Appl. Phys. 40 (2007) 7593-7600.
\bibitem{Szal08}L. Szalm$\acute{a}$s, Variable slip coefficient in binary lattice Boltzmann models, Cent. Eur. J. Phys. 6 (2008) 786-791.
\bibitem{Guo2009} Z.L. Guo, P. Asinari, C.G. Zheng, Lattice Boltzmann equation for microscale gas flows of binary mixtures, Phys. Rev. E. 79 (2009) 026702.
\bibitem{Lim2002} C.Y. Lim, C. Shu, X.D. Niu, Y.T. Chew, Application of lattice Boltzmann method to simulate microchannel flows, Phys. Fluids. 14 (2002) 2299-2308.
\bibitem{Xu2013} Z.M. Xu, Z.L. Guo, Pressure Distribution of the Gaseous Flow in Microchannel: A Lattice Boltzmann Study, Commun. Comput. Phys. 14 (2013) 1058-1072.
\bibitem{He97} X. He, L.-S. Luo, A priori derivation of the lattice Boltzmann equation, Phys. Rev. E. 55 (1997) 6333-6336.
\bibitem{Abe97} T. Abe, Derivation of the lattice Boltzmann method by means of the discrete ordinate method for the Boltzmann equation, J. Comput. Phys. 131 (1997) 2410-246.
\bibitem{ASina08} P. Asinari, L.-S. Luo, A consistent lattice Boltzmann equation with baroclinic coupling for mixtures, J. Comput. Phys. 227 (1997) 3878-3895.
\bibitem{Guo2006} Z.L. Guo, T.S. Zhao,Y. Shi, Physical symmetry, spatial accuracy, and relaxation time of the lattice Boltzmann equation for microgas flows, J. Appl. Phys. 99 (2006) 074903.
\bibitem{Guo2008} Z.L. Guo, C.G. Zheng, B.C. Shi, Lattice Boltzmann equation with multiple effective relaxation times for gaseous microscale flow, Phys. Rev. E. 77 (2008) 036707.
\bibitem{Cercignani90}C. Cercignani, Mathematical Methods in Kinetic Theory, Plenum Press, New York, 1990.
\bibitem{Yalamov94}Y.I. Yalamov, A.A. Yushkanov, S.A. Savkov, Boundary conditions for the slippage of a binary mixture of gases and their application in the dynamics of aerosols. I. Flow of a mixture of gases along a solid plane wall, J. Eng. Phys. Thermophys. 66 (1994) 421-426.
\bibitem{Sharipov02}F. Sharipov, D. Kalempa, Gaseous mixture flow through a long tube at arbitrary Knudsen numbers, J. Vac. Sci. Technol. A 20 (2002) 814-822.
\bibitem{Tang2004} G.H. Tang, W.Q. Tao, Y.L. He, Lattice Boltzmann method for simulating gas flow in microchannels, Int. J. Mod. Phys. C. 15 (2004) 335-347.
\bibitem{Succi2005} M. Sbragaglia, S. Succi, Analytical calculation of slip flow in lattice Boltzmann models with kinetic boundary conditions, Phys. Fluids. 17 (2005) 093602.
\bibitem{Li2009} W.Z. Li, W.N. Zhou, Lattice Boltzmann simulation of micro Poiseuille flow in curved rough channels, MNHMT (2009) 18222.
\bibitem{Guo2011} Z.L. Guo, B.C. Shi, C.G. Zheng, Velocity inversion of micro cylindrical Couette flow: A lattice Boltzmann study, Comput. Math. Appl. 61 (2011) 3519-3527.
\bibitem{Ivchenko1997} I.N. Ivchenko, S.K. Loyalka, R.V. Tompson, Slip coefficients for binary gas mixtures, J. Vac. Sci. Technol. A 15 (1997) 2375-2381.
\end{thebibliography}
\end{document}